\documentclass[STIX1COL]{WileyNJD-v2}
\usepackage{moreverb}
\usepackage{amssymb,amsmath,relsize,enumerate,bbm,graphicx,subcaption,caption}

\usepackage{natbib}
\usepackage{multirow}
\usepackage{colortbl}
\usepackage{pdflscape}

\usepackage{tikz}
\usetikzlibrary{shapes.geometric, arrows}
\tikzstyle{box} = [rectangle, minimum width=3cm, minimum height=1cm, text centered, draw=black]
\tikzstyle{arrow} = [thick,->,>=stealth]

\newcommand\BibTeX{{\rmfamily B\kern-.05em \textsc{i\kern-.025em b}\kern-.08em
T\kern-.1667em\lower.7ex\hbox{E}\kern-.125emX}}

\articletype{Article Type}%

\received{<day> <Month>, <year>}
\revised{<day> <Month>, <year>}
\accepted{<day> <Month>, <year>}

\newcommand{\E}{\mathbb{E}}

\newcommand{\var}{\mbox{Var}}

\def\*#1{\boldsymbol{#1}}
\def\~#1{{\cal #1}}

\begin{document}

\title{Estimating risk factors for pathogenic dose accrual from longitudinal data}

\author[1]{Daniel K. Sewell*}

\author[2,3]{Kelly K. Baker}

\authormark{Sewell \& Baker}

\address[1]{\orgdiv{Department of Biostatistics}, \orgname{University of Iowa}, \orgaddress{\state{Iowa}, \country{USA}}}

\address[2]{\orgdiv{Department of Occupational and Environmental Health}, \orgname{University of Iowa}, \orgaddress{\state{Iowa}, \country{USA}}}

\corres{*Daniel K. Sewell, 145 N. Riverside Dr., Iowa City, IA 52242, USA. \email{daniel-sewell@uiowa.edu}}

\presentaddress{145 N. Riverside Dr., Iowa City, IA 52242, USA.}

\abstract[Abstract]{Estimating risk factors for incidence of a disease is crucial for understanding its etiology.  For diseases caused by enteric pathogens,  off-the-shelf statistical model-based approaches do not consider the biological mechanisms through which infection occurs and thus can only be used to make comparatively weak statements about association between risk factors and incidence.  Building off of established work in quantitative microbiological risk assessment, we propose a new approach to determining the association between risk factors and dose accrual rates. Our more mechanistic approach achieves a higher degree of biological plausibility, incorporates currently-ignored sources of variability, and provides regression parameters that are easily interpretable as the dose accrual rate ratio due to changes in the risk factors under study.  We also describe a method for leveraging information across multiple pathogens. The proposed methods are available as an R package at \url{https://github.com/dksewell/dare}.  Our simulation study shows unacceptable coverage rates from generalized linear models, while the proposed approach empirically maintains the nominal rate even when the model is misspecified.  Finally, we demonstrated our proposed approach by applying our method to infant data obtained through the PATHOME study (\url{https://reporter.nih.gov/project-details/10227256}), discovering the impact of various environmental factors on infant enteric infections. }


\jnlcitation{\cname{%
\author{D.K. Sewell}, and
\author{Kelly K. Baker}} (\cyear{2024}), 
\ctitle{Estimating risk factors for dose accrual from longitudinal data}, \cjournal{TBD}, \cvol{202x;00:0--0}.}

\maketitle

\section{Introduction}
\label{sec:intro}
Enteric infections are a significant source of morbidity and mortality globally.  The World Health Organization estimates that each year there are nearly 1.7 billion cases of childhood diarrheal disease, with over 440,000 deaths in children under five years old \citep{whoDiarrhea}.  While this disproportionately impacts low- to middle-income countries, high-income countries, too, are highly impacted.  For example, \textit{Clostridioides difficile} infections alone affect roughly 500,000 individuals in the United States each year, leading to around 30,000 deaths \citep{feuerstadt2023}.  Obtaining a deeper understanding of transmission dynamics and how they vary according to individual-level characteristics is critical to understanding disease etiology \citep{ward2013estimating}, which in turn leads to more informed intervention design and health policies.

\subsection{Incidence estimation}
Incidence is one of the most core quantitative epidemiological measures available to help understand infectious disease transmission dynamics \citep{bruce2008quantitative}. The simplest approach to describing infection rates is through the incidence rate, defined to be the number of new cases over a specified time interval, or, relatedly, through the incidence proportion given by the proportion of at-risk individuals to become infected over a specified time interval.  Ascertaining how incidence varies in subpopulations associated with specific risk factors in this way is predicated on several factors, including a clear delineation of the population into subpopulations, and population level surveillance.  While the latter issue can be addressed through sampling techniques and established statistical inference procedures, the former leads to intractability for multiple risk factors, as this number grows exponentially with the number of individual-level factors.  These two problems- the requirement to sample and having a large number of partitions of the population- synergize in that within each subpopulation, sufficient numbers of subjects must be recruited so as to reliably estimate the rate at which new cases occur.  This is, of course, yet further exacerbated by rare diseases.  To ameliorate these issues, generalized linear models (GLMs) based on the Poisson distribution has been used, which makes the additional assumption that the the variation in incidence across subpopulations can (on the log scale) be described mathematically as an additive combination of the subpopulations' risk factors \citep{frome1985use}.

An additional problem with the above approaches arises when one or more individual-level risk factor is not categorical or finite in nature.  In such a case either arbitrary thresholding must occur, or some other model-based approach is taken. Several such approaches can be found in the extant literature. Some researchers turn to logistic regression \citep[e.g.,][]{cohen2022sarscov2}, estimating the change in odds of becoming newly infected over a specified time period due to individual-level risk factors.  However, this approach requires each individual recruited to the study be observed for the same amount of time.  In some cases this is feasible, but this is often not the case due to a variety of reasons (for example, if study participants schedule their follow-up time within a varying time window from baseline, or if it is not possible to precisely schedule biological specimen collection).  Another approach is the use of survival models, such as the Cox Proportional Hazards model \citep[e.g.,][]{schwarz2008placental}. This approach, however, typically requires that the new infection causes acute symptoms which allow the infection to be surveilled passively, and that the incubation period is either known or negligible.  Another approach which is sufficiently flexible to handle varying observation lengths is the use of the complementary log-log link function in a GLM based on the Bernoulli distribution \citep[e.g.,][]{{verburgh2021similar}}.  By using the log of the individuals' time intervals as an offset in the model, one can estimate the rate at which new cases develop for a given set of covariates.

\subsection{A mechanistic view}
\begin{figure}
    \centering
    \includegraphics[width=0.9\textwidth]{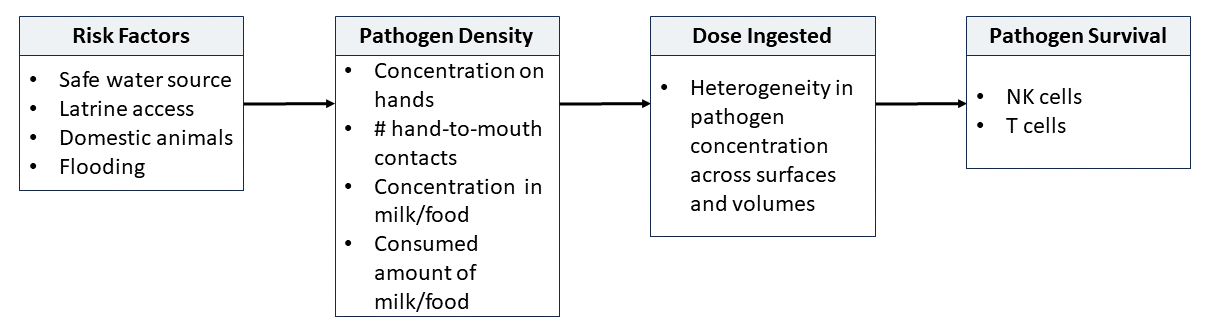}
    \caption{Sources of variability in potential infections, along with examples of influential factors for each.}
    \label{fig:sources_of_variability}
\end{figure}

There are several key steps by which microorganisms in the environment result in infecting an individual, each of which is an important source of variability. First, environmental and behavioral risk factors impact what an individual is exposed to. As a running example, consider access or lack of access a child has to a private latrine. Second, these risk factors lead to varying dose density of the various fomites, vectors, and vehicles to which an individual is exposed, and combined with stochastic behaviors, such as quantity of media consumed or number of hand-to-mouth contacts, result in highly variable expected ingested dose. E.g., even after conditioning on latrine access, the dose density of an enteric pathogen on a child's hands will vary stochastically based on who else has used the latrine recently, and what the child happens to touch and how often on a given day.  A third source of variability is in the actual dose received given the expected ingested dose. E.g., conditioning on the dose density on a child's hands and a certain number of hand-to-mouth contacts, the actual dose ingested will still vary due to heterogeneity of pathogen concentration on the hands' surface, which part of the hand is in contact with which part of the mouth and for how long. Additionally, temporal pathogen-environment interactions can enhance or reduce pathogen survival in the environment, affecting viability of  pathogen dose.  Finally, pathogens have varying abilities to propagate within a host \citep{parker2018hostmicrobe}.  Infection ``can be described by competing processes of birth and death within the host, infection resulting when birth is sufficient to produce a body burden above some critical level to induce the effect'' \citep[][p. 268]{haas2014quantitative}.  That is, conditioning on the dose ingested, each pathogen has a random chance at surviving to be able to cause an infection, affected by, e.g., a host's NK cells or T cells.  These last two criteria are, according to \cite{haas2014quantitative}, required to achieve biological plausibility. Examples of these sources of variability are summarized graphically in Figure \ref{fig:sources_of_variability}.  Together, these steps result in highly varying dose accrual rates across individuals. 

\subsection{Contributions of this paper}
We make the following two arguments.  First, as incidence is fundamentally about describing infection rates in populations or sub-populations, it fails to address the mechanisms by which individuals become infected.  In incidence models which examine associations with risk factors, only the first source of variability listed above, that of the risk factors, is addressed.  In the language of \cite{haas2014quantitative}, such incidence models fail to achieve biological plausibility as an infection model.  If the mechanisms of infection are the quantities of interest, than dose accrual rates, rather than incidence rates, ought to be the focus of one's analysis. 

Second, stating that a unit increase in $x$ leads to a $\eta$ increase in the incidence rate is a fundamentally weaker statement than stating that a unit increase in $x$ leads to a $\eta$ increase in the pathogenic dose ingested for a given dose-response model.  This claim is based on the fact that having knowledge about the effect a risk factor has on the dose accrual rate allows one to determine the effect on the incidence rate, whereas the reverse is not true.

To address these two points, we propose a novel model for assessing the impact of individual level features on the dose accrual rates for the pathogen under study. Our proposed approach, which we call Dose Accrual Risk Estimation, or \textit{DARE}, has the following features.
\begin{itemize}
    \item By building on Quantitative Microbial Risk Assessment (QMRA) techniques, we both satisfy the plausibility criteria given by \cite{haas2014quantitative} and account for the salient sources of variability listed above.
    \item Rather than focusing on incidence estimation, our approach directly estimates the rate at which pathogenic dose is accrued per time unit.
    \item Our approach can handle varying time intervals between individuals' repeated measurements.
    \item We further provide a method for leveraging information across the analyses of multiple pathogens that uses recent work on linear subspace shrinkage techniques.
    \item Our DARE methodology is available through the R package \textit{dare}, available through github.
\end{itemize}

The remainder of this paper is as follows.  Section \ref{sec:methods} describes our proposed longitudinal model with its derivation, along with an approach for leveraging information across the analyses of multiple pathogens.  Section \ref{sec:simstudy} describes a simulation study analyzing the estimation performance of our approach.  Section \ref{sec:pathome} illustrates our proposed approach based on data collected through the PATHOME study \citep{baker2023protocol}.  Finally, we provide a discussion in Section \ref{sec:discussion}.

\section{Methods}
\label{sec:methods}

\subsection{A dose accrual model}
\label{subsec:dare}
Suppose our study involves $N$ at-risk individuals, and for the $i^{th}$ individual we observe them at the end of each of $T_i$ intervals before either an infection is detected or they exit the study.  We will denote the length of these time intervals as $\tau_{it}$ and the binary outcome as $y_{it}$, where $y_{it}$ equals one if after a period of $\tau_{it}$ they are infected and zero otherwise.  
For an infection to occur during an exposure window, a subject must ingest one or more infection-causing pathogens. In addition, for an infection to occur, one or more of these ingested pathogens must survive within the host long enough to begin the infection.  Based on variations of these two processes, a plethora of dose-response models have been developed (see, e.g., \cite{haas2014quantitative}).  Our proposed approach is agnostic to the specific dose-response model, in that what we propose should be compatible with any such model.  We will therefore denote the dose-response model as $P_{\theta}(\cdot)$, where $\theta$ is the set of associated dose-response parameters.

The two most common dose-response models are the exponential model and the beta-Poisson model \citep{soller2006use}.  The only parameter of the exponential dose-response model is the (iid) survival rate of the organisms, and the model itself is given by
\begin{equation}
    \label{eq:exponential}
    P_\theta(D) := 1 - e^{-\theta D},
\end{equation}
where $D$ is the expected dose.  The beta-Poisson model is predicated on host variability leading to different survival rates.  The beta-Poisson model is then parameterized by, as its name suggests, the two shape parameters of the beta distribution describing the between-host distribution of organism survival probabilities.  Thanks to theoretical work in \cite{furumoto1967mathematical}, the beta-Poisson dose-response model is commonly approximated as
\begin{equation}
    \label{eq:betapoisson}
    P_\theta(D) := 1 - \left( 1 + \frac{D}{\theta_2} \right)^{-\theta_1}.
\end{equation}

If we were to know the mean dose $D_{it}$ of the $t^{th}$ exposure period for subject $i$, the conditional likelihood of our data would be given as
\begin{equation}
    \Pr(y_{1,1},\ldots,y_{N,T_N}| D_{1,1},\ldots,D_{N,T_N},\theta) 
    =
    \prod_{i=1}^N \prod_{t=1}^{T_i} \big[P_\theta(D_{it})\big]^{y_{it}} \big[1 - P_\theta(D_{it})\big]^{1 - y_{it}}.
    \label{eq:cond_lik}
\end{equation}
In highly controlled experiments, $D_{it}$ would typically represent the dose density in a given medium being ingested.  In observational studies, however, $D_{it}$ in Eq. (\ref{eq:cond_lik}) represents an agglomeration of pathways.  
That is, $D_{it}$ can be thought of as an 
\textit{expected ingested dose}
that encompasses the 
collection of dose density on hands, in water, in milk, in food, etc., as well as behaviors that lead to ingestion such as hand-to-mouth contacts and quantity of media consumed.  $D_{it}$ should be considered stochastic for dose densities and time interval-specific behaviors will not be the same for two individuals with the same risk factors, nor even the same individual at different times. 

Historically, expected pathogen dose has been modeled using log-normal distributions \citep{haas2014quantitative}, and we do not break from this tradition here.  Letting $X_{it}$ denote a $1\times J$ vector of risk factors for individual $i$ during the $t^{th}$ time interval, the expected ingested dose, $D_{it}$, is a stochastic quantity depending on $X_{it}$ (while the actual dose ingested is a separate stochastic quantity 
captured in the dose-response model $P_\theta$). Specifically, we model $D_{it}$ as follows:
\begin{equation}
    \label{eq:lognormal}
    D_{it} \sim \ell N(X_{it}\*\beta + \log(\tau_{it}),\sigma^2),
\end{equation}
where $\ell N(\mu,\sigma^2)$ is the log-normal distribution with location parameter $\mu$ and log-scale parameter $\sigma$, and $\beta:=(\beta_1,\ldots,\beta_J)$ is the vector of log-rate regression coefficients.   Eq. (\ref{eq:lognormal}) implies that the rate of accrual of expected dose is 
\begin{equation}
    \label{eq:accrual_rate}
    e^{X_{it}'\beta + \frac{\sigma^2}{2}},
\end{equation}
leading to the first and second central moments of the expected dose over a time interval of length $\tau_{it}$ to be
\begin{align*}
    \E(D_{it} | X_{it},\beta) 
    & = 
    \tau_{it}e^{X_{it}'\beta + \frac{\sigma^2}{2}},
    & &
    \var(D_{it} | X_{it},\beta)
    = 
    \E^2(D_{it} | X_{it},\beta) \left( e^{\sigma^2} - 1 \right).
\end{align*}
\noindent One note of interest is that if the exponential dose-response model is used, as $\sigma\to0$ we obtain the GLM based on the binomial distribution with the complementary log-log link function.

In summary, the two dose-response models described above assume that the number of pathogens ingested follows a Poisson distribution given the mean dose.  Each pathogen ingested is assumed to either have a constant survival probability (exponential dose-response model) or host-specific survival probability (beta-Poisson dose-response model). The expected ingested dose itself, due to variability in the environment and host behavior, follows a log-normal distribution, and this distribution depends on observable risk factors.  In particular, the rate at which dose is accrued changes by a factor of $e^{\beta_j}$ from a unit increase in the $j^{th}$ covariate, keeping all other covariates the same.  Together, we obtain the unconditional likelihood of our data as
\begin{align}
    \nonumber
     &\Pr(y_{1,1},\ldots,y_{N,T_N}| \theta,\beta,\sigma^2) 
     &\\ 
    \label{eq:dare}
     & =
    \prod_{i=1}^N \prod_{t=1}^{T_i} 
    \int_{-\infty}^\infty \left[P_\theta\left(\tau_{it}e^{X_{it}'\beta + \sigma z_{it}}\right)\right]^{y_{it}} \left[1 - P_\theta(\tau_{it}e^{X_{it}'\beta + \sigma z_{it}})\right]^{1 - y_{it}}\phi(z_{it})dz_{it}, &
\end{align}
where $\phi(\cdot)$ is the standard normal probability density function. We will refer to Eq. (\ref{eq:dare}) as the DARE model.  Note that in computing the DARE likelihood, the univariate integrals can be solved using standard numerical integration methods, such as Gaussian quadrature. 

In the DARE model, there is an important issue of identifiability that must be acknowledged.  In both the exponential and beta-Poisson dose-response models we have perfect confounding involving the intercept term.  That is, $\beta_1$ (assuming $X_{it1}=1\forall i,t$) is perfectly confounded with $\theta$ from the exponential model and $\theta_2$ in the beta-Poisson model.  We therefore fix the $\theta$ or $\theta_2$ to be 1, and estimate $\beta_1$ as an unconstrained (and uninterpretable) parameter.  However, this issue further necessitates some caution in interpretation of any modeling results.  Were we to have modeled $\log(\theta)$ in the exponential model or $-\log(\theta_2)$ in the beta-Poisson model\footnote{Note that as $-\log(\theta_2)$ increases, or equivalently $\theta_2$ decreases, the survival probability of each organism increases.} as a linear combination of our covariate vector $X_{it}$, the corresponding regression coefficients would again be perfectly confounded with $\beta$.  Hence as one interprets any analysis output using the DARE model, one is bound to determine using context and domain expertise whether the effect of a specific covariate is on the dose accrual rate or the survival rate of the pathogens.  

\subsection{Combining results from multiple pathogens}
\label{subsec:subset}
We now expand our discussion to include contexts where we are measuring multiple pathogens. It will often be the case that a covariate will act on the rate of dose accrual similarly between certain pathogens if, for example, two pathogens are often transmitted through the same vector, fomite, or vehicle.  Yet a hard constraint setting these regression coefficients to be equal is highly implausible.  For example, if one pathogen is solely waterborne, while another pathogen is both waterborne and transmitted through food, both of these pathogens' rate of accrual will change similarly with respect to safe water access, yet clearly there will still be important differences; the former pathogen will be successfully mitigated through a water intervention, while the latter has a minimum threshold of effect that cannot reach disease elimination solely through such a water intervention.  In other words, we wish to shrink certain regression coefficients towards each other in a data driven way without imposing unrealistic hard constraints of equality. The recent SUBSET method of \cite{sewell2024posterior} provides tools to accomplish this.  The idea is to find a linear subspace to shrink towards, and use exponential tilting of the prior to induce the desired shrinkage.  Unlike most statistical shrinkage methods which focus on point estimation, this approach shrinks the entire posterior, thereby influencing all resulting inference.  By adapting SUBSET to our present context as described below, we are able to leverage information across pathogen-specific analyses in a data driven way that, while not imposing any equality constraints, allows the data to dictate the degree to which certain parameters ought to be similar across pathogens.

Let $\beta_{(k)j}$ denote the $j^{th}$ regression coefficient for pathogen $k$, $k=1,\ldots,K$, and similarly let $\theta_{(k)}$ and $\sigma^2_{(k)}$ be the dose-response model parameter(s) and dose variance for the $k^{th}$ pathogen respectively. The entire parameter vector of regression coefficients, dose response parameters, and dose variance is of length $Q := KJ'$, where $J':=(J+ |\theta| + 1 )$, and we will denote it as
\[
    \eta:=(\beta_{(1)1},\ldots,\beta_{(1)J},\theta_{(1)},\sigma^2_{(1)},\beta_{(2)1},\ldots,\beta_{(K)J},\theta_{(K)},\sigma^2_{(K)}).
\]
 Let $L$ denote a matrix whose $Q$ rows represent the unknown parameters and whose columns dictate which parameters are free and which have an equality constraint.  We can construct $L$ in the following manner, assuming that the $K$ pathogens' intercept, $\theta$, and $\sigma^2$ will not be shrunk towards each other. For a set ${\cal S}\subseteq [K]$, let $v_M({\cal S})$ denote the $M\times 1$ column vector such that its $m^{th}$ element equals 1 if $m\in{\cal S}$ and 0 otherwise; and let $I(j,{\cal S})$ denote the $Q\times |{\cal S}|$ matrix equal to the $Q\times Q$ identity matrix, selecting those $|{\cal S}|$ columns corresponding to $\{ J'(k - 1) + j: k\in{\cal S} \}$.  Finally, for $j=1,\ldots,J$ let ${\cal S}_j\subseteq[1:K]$ denote the set of pathogens whose $j^{th}$ coefficients we wish to shrink towards each other, and ${\cal S}_j^C$ denote its complement\footnote{Note that ${\cal S}_j$ could be the empty set, in which case no shrinkage is performed on the $j^{th}$ regression coefficients.}.  Then we can set $L$ to be 
 \begin{align}
    \nonumber
     L :=
     & 
     \Big(
        I(1,[K]), I(2,{\cal S}_2^C),v_K({\cal S}_2)\otimes v_{J'}(\{2\}), 
    &\\
    & \ldots,I(J,{\cal S}_J^C),v_K({\cal S}_J)\otimes v_{J'}({J}),
        I(J+1,[K]),\ldots,I(J',[K])
     \Big).
 \end{align}
The linear subspace we wish to shrink towards is $span(L)$.  Only the regression coefficients that have a $S_j\neq \emptyset$ experience shrinkage.  

As an example, consider the matrix image of $L$ given in Figure \ref{fig:subset_illustration_of_L}, where there are four pathogens; three covariates consisting of the intercept, $X_1$, and $X_2$; the intercepts are not shrunk towards each other; all $X_1$ coefficients are shrunk towards each other; and the $X_2$ coefficients are shrunk together for only pathogens 2 and 4.  That is, ${\cal S}_1=\emptyset$, ${\cal S}_2=\{1,2,3,4\}$, ${\cal S}_3 = \{2,4\}$ (and $\sigma^2$ and $\alpha$ do not experience shrinkage).

\begin{figure}
    \centering
        \begin{tabular}{c c |p{0.2cm}p{0.2cm}p{0.2cm}p{0.2cm}|p{0.2cm}|p{0.2cm}p{0.2cm}p{0.2cm}|p{0.2cm}p{0.2cm}p{0.2cm}p{0.2cm}|p{0.2cm}p{0.2cm}p{0.2cm}p{0.2cm}|}
            \multicolumn{2}{c}{} &
            \multicolumn{16}{p{3.2cm}}{$\overbrace{{\color{white}aaaaaaaaaaaaaaaaaaaaaaaaaaaaaaaaaaaaaaaaaaaaa}}^{\mbox{Pathogen set to shrink}}$} \\
            \multicolumn{1}{c}{} & \multicolumn{1}{c}{} & 
             \multicolumn{2}{p{0.4cm}}{} & \multicolumn{2}{p{0.4cm}}{\hspace{-0.5pc}$\emptyset$} &
            \multicolumn{1}{p{0.2cm}}{\rotatebox{45}{\hspace{-0.5pc}$\{1,2,3,4\}$}} & 
            \multicolumn{1}{p{0.2cm}}{} & \multicolumn{2}{p{0.6cm}}{\rotatebox{45}{$\{2,4\}$}} & 
            \multicolumn{2}{p{0.4cm}}{} & \multicolumn{2}{p{0.4cm}}{\hspace{-0.5pc}$\emptyset$} &
            \multicolumn{2}{p{0.4cm}}{} & \multicolumn{2}{p{0.4cm}}{\hspace{-0.5pc}$\emptyset$} \\ 
            \hline
            \multirow{5}{*}{Pathogen 1}& Intercept & \cellcolor{black} & & & & & & & & & & & & & & & \\
            & $X_1$ & & & & & \cellcolor{black} & & & & & & & & & & & \\
            & $X_2$ & & & & & & \cellcolor{black} & & & & & & & & & & \\
            & $\sigma^2$ & & & & & & & & & \cellcolor{black} & & & & & & & \\
            & $\alpha$ & & & & & & & & & & & & & \cellcolor{black} & & & \\
            \hline
            \multirow{5}{*}{Pathogen 2}& Intercept & & \cellcolor{black} & & & & & & & & & & & & & & \\
            & $X_1$& & & & & \cellcolor{black} & & & & & & & & & & & \\
            & $X_2$ & & & & & & & & \cellcolor{black} & & & & & & & & \\
            & $\sigma^2$ & & & & & & & & & & \cellcolor{black} & & & & & & \\
            & $\alpha$ & & & & & & & & & & & & & & \cellcolor{black} & & \\
            \hline
            \multirow{5}{*}{Pathogen 3}& Intercept & & & \cellcolor{black} & & & & & & & & & & & & & \\
            & $X_1$& & & & & \cellcolor{black} & & & & & & & & & & & \\
            & $X_2$ & & & & & & & \cellcolor{black} & & & & & & & & & \\
            & $\sigma^2$ & & & & & & & & & & & \cellcolor{black} & & & & & \\
            & $\alpha$ & & & & & & & & & & & & & & & \cellcolor{black} & \\
            \hline
            \multirow{5}{*}{Pathogen 4}& Intercept & & & & \cellcolor{black} & & & & & & & & & & & & \\
            & $X_1$& & & & & \cellcolor{black} & & & & & & & & & & & \\
            & $X_2$ & & & & & & & & \cellcolor{black} & & & & & & & & \\
            & $\sigma^2$ & & & & & & & & & & & & \cellcolor{black} & & & & \\
            & $\alpha$ & & & & & & & & & & & & & & & & \cellcolor{black} \\
            \hline
            \multicolumn{1}{c}{} & \multicolumn{1}{c}{} &
            \multicolumn{4}{p{0.8cm}}{Intercept} & 
            \multicolumn{1}{p{0.2cm}}{$X_1$} &  
            \multicolumn{1}{p{0.2cm}}{} & \multicolumn{2}{p{0.6cm}}{$X_2$} &
            \multicolumn{2}{p{0.4cm}}{} & \multicolumn{2}{p{0.4cm}}{$\sigma^2$} &
            \multicolumn{2}{p{0.4cm}}{} & \multicolumn{2}{p{0.4cm}}{$\alpha$} \\
            \multicolumn{2}{c}{} &
            \multicolumn{16}{p{3.2cm}}{$\underbrace{{\color{white}aaaaaaaaaaaaaaaaaaaaaaaaaaaaaaaaaaaaaaaaaaaaa}}_{\mbox{Parameter to be constrained}}$}
        \end{tabular}
    \caption{Illustration of the matrix $L$ to shrink certain parameters towards equality for certain pathogens.  The intercept, $\sigma^2$, and $\alpha$ are not shrunk; the coefficients for $X_1$ are shrunk towards each other; and the coefficients for $X_2$ are shrunk towards each other for pathogens 2 and 4 only. The subsets ${\cal S}_j$ are labeled above, while the parameters being constrained are labeled below.}
    \label{fig:subset_illustration_of_L}
\end{figure}

The SUBSET prior multiplicatively changes the joint prior on $\eta$ by a factor of 
\begin{equation}
   \exp\left\{
        -\frac\nu2\eta'(I_Q - L(L'L)^{-1}L')\eta,
   \right\}
\end{equation}
which effectively penalizes areas of the parameter space distant from the linear subspace, where $I_Q$ is the $Q\times Q$ identity matrix, and $\nu$ is a positive valued scalar that determines the level of shrinkage.  The value of $\nu$ can be selected in a data-driven way by maximizing the Bayes factor \citep[see][for details]{sewell2024posterior}.  If after analyzing the pathogens separately we denote the mode and the hessian of the negative log posterior for $\eta$ as $m_n$ and $\Omega_n$ respectively, then the large sample approximation of the joint posterior of $\eta$ under the SUBSET prior induced by $L$ is given by 
\begin{align}\nonumber
    \eta|\mbox{data} & \sim N\big(
        \tilde m_n, \widetilde \Omega_n^{-1}
    \big), 
    &\\ \nonumber
    \mbox{where } \widetilde \Omega_n & := \Omega_n + \nu(I_Q - L(L'L)^{-1}L'),
    &\\
    \tilde m_n & := \widetilde \Omega_n^{-1}\Omega_nm_n.
\end{align}

\section{Simulation study}
\label{sec:simstudy}
We wished to see how well we could estimate the unknown regression coefficients of the DARE model using the beta-Poisson dose-response model, as well as determine how well the complementary log-log binomial GLM- the closest existing model to DARE- can recovery the true parameter values. The true values of $\beta$ in our simulation study were $(-4.6,0,0.5,1)$ corresponding to an intercept and three covariates, each of which were randomly drawn for each individual from a standard normal distribution.  We generated data according to both the exponential and the beta-Poisson.  We considered all combinations of $\sigma\in \{1,2,3\}$ and- for the beta-Poisson model- $\theta_1\in\{1,2,3\}$.  These values, excluding the non-intercept regression coefficients, were selected to resemble the average values of those estimated from the PATHOME data described in Section \ref{sec:pathome}, for which the average intercept (taken over all pathogens and age groups) was -4.6, the average $\sigma$ was 3.0, and the average $\theta_2$ was 2.5.  

To evaluate our approach, we computed the coverage rates of the 95\% credible intervals and examined the estimated regression coefficients.  For each dose-response model we simulated 1000 data sets.  Each data set included 215 subjects each measured at times 1, 3, 5, 7, and 14 or until an infection was detected; again, these values were selected to mimick the PATHOME study described in Section \ref{sec:pathome}.

For both the DARE model and the complementary log-log GLM, we used weakly regularizing priors.  Specifically, we used a $N(0,2.5^2)$ prior for the regression coefficients with the exception of the intercept for which we used a $N(0,10^2)$ prior.  For the DARE model fits, we used a gamma prior with shape and rate both equal to 2 for $\sigma$ and an exponential with mean 1 for $\theta_1$.  

\begin{figure}
    \centering
    \includegraphics[width=0.8\linewidth]{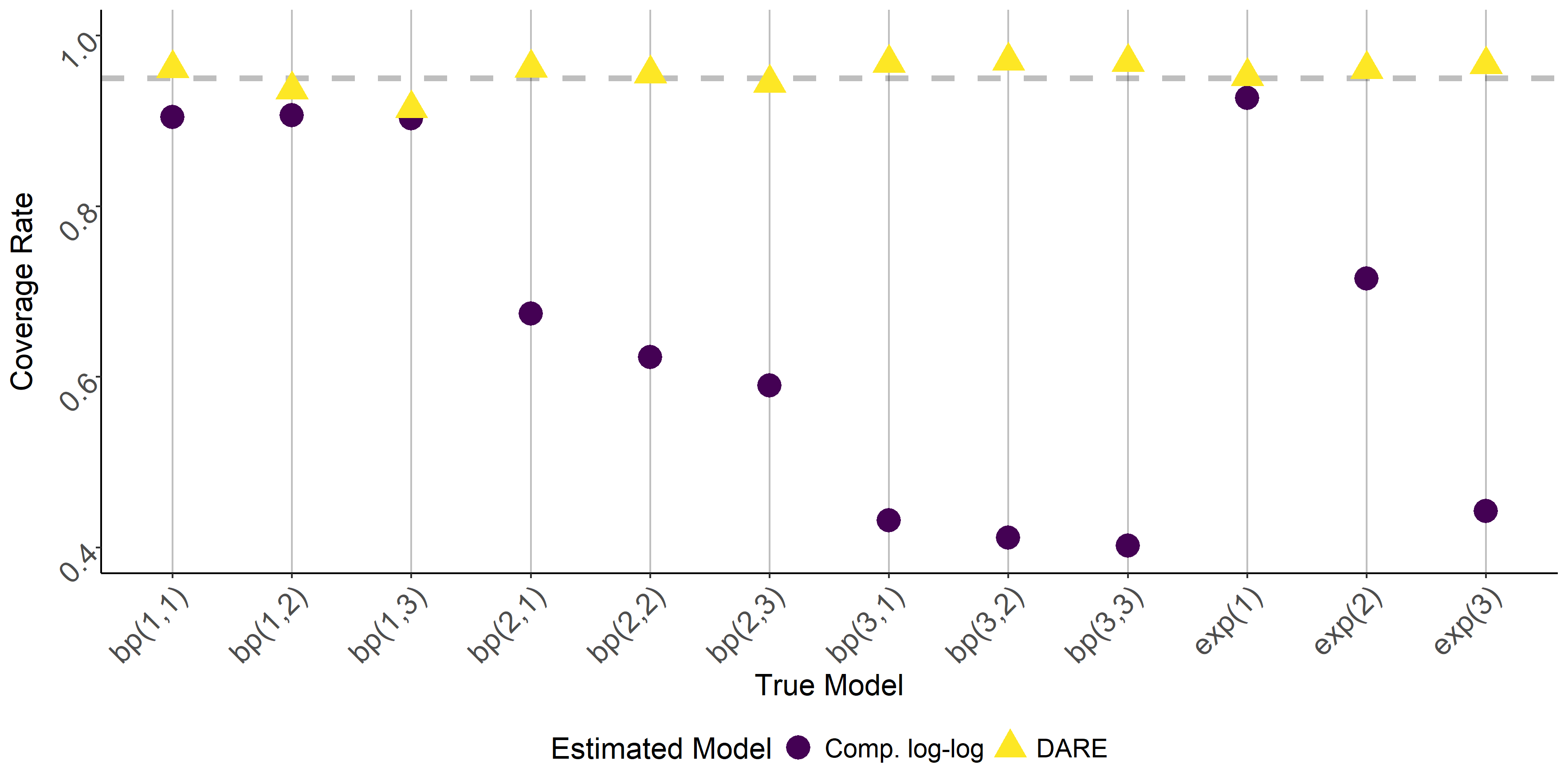}
    \caption{Simulation study results for the coverage rate of 95\% credible intervals, comparing a GLM with the complementary log-log link with DARE based on the beta-Poisson dose-response model.  The true model is given in the form $\text{bp}(\sigma,\theta_1)$ or $\text{exp}(\sigma)$ for the beta-Poisson and exponential dose-response models respectively. The nominal rate (0.95) is given in the dashed gray line.}
    \label{fig:simstudy-coverage}
\end{figure}

Figure \ref{fig:simstudy-coverage} shows the coverage rates for the GLM as well as the DARE model. When $\sigma=1$, the GLM has lower than nominal but respectable coverage rate.  However, for larger values of $\sigma$ the coverage becomes unacceptably low.  As $\theta_1$ increases, the coverage rate decreases, but not as severely as with increases in $\sigma$.  Meanwhile, the beta-Poisson DARE model appears to maintain the nominal coverage rate, even when the dose-response model is misspecified.

\begin{figure}
    \centering
    \includegraphics[width=\linewidth]{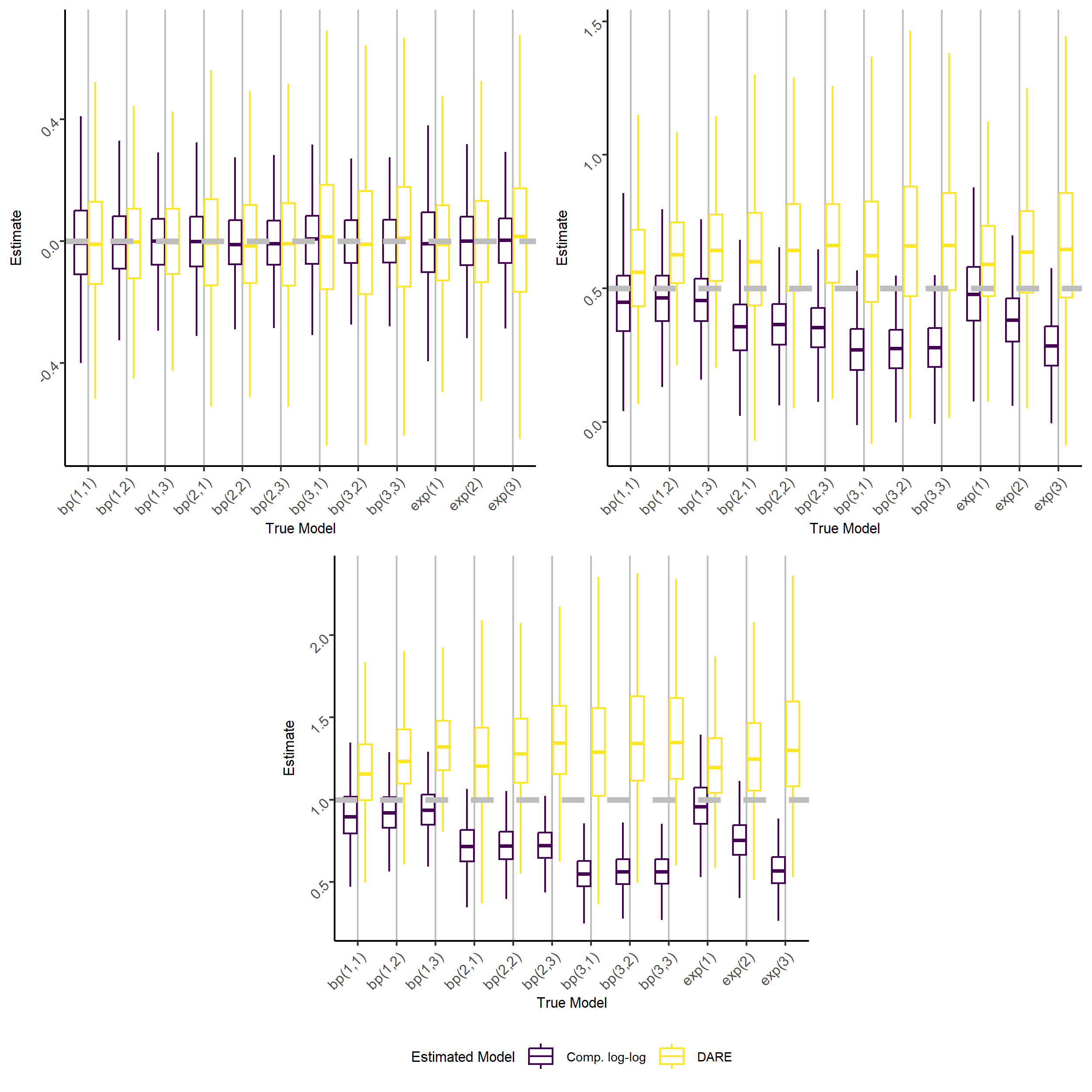}
    \caption{Simulation study results for the point estimates of the three regression coefficients ($\beta_1 = 0$, $\beta_2 = 0.5$, and $\beta_3 = 1$), comparing a GLM with the complementary log-log link with DARE based on the beta-Poisson dose-response model.  The true model is given in the form $\text{bp}(\sigma,\theta_1)$ or $\text{exp}(\sigma)$ for the beta-Poisson and exponential dose-response models respectively.  The true values are given by the dashed gray lines.}
    \label{fig:simstudy-estimates}
\end{figure}

Figure \ref{fig:simstudy-estimates} graphically displays the estimated regression coefficients from both the DARE and GLM model fits. From this we see that the GLM estimates tend to be negatively biased for non-zero coefficients while the DARE estimates tend towards a positive bias.  However, while the DARE bias appears relatively stable across values of $\sigma$ and $\theta_1$, the negative bias of the GLM gets progressively worse as $\sigma$ increases.

\section{Enteric infections in infants}
\label{sec:pathome}
The Pathogen Transmission and Health Outcome Models of Enteric Disease (PATHOME) study aims to use a One Health approach to better understand enteric pathogen transmission in low- to middle-income countries.  Infants from 0-12 months old and their households were recruited into the PATHOME study from low-income and middle-income neighborhoods from Nairobi and Kisumu, Kenya.  In each city, we selected households from both low and middle income neighborhoods. In total, we obtained microbiological data on 214 infants.

On the first day of participating in the study, caregivers were given a survey on socioeconomic conditions, behaviors, and household health.  While other data elements, such as geotracking animals, were collected as part of the PATHOME study, they will not be our focus here.  The variables we analyzed included city, socioeconomic status of the neighborhood (SES), whether the household's compound flooded, whether the household owned domestic animals\footnote{This included chickens, ducks, pigs, goats, sheep, cows, dogs, rabbits, donkeys, and turkeys.}, whether animals not owned by the household entered their compound, and whether the family had access to a private latrine. These data elements are summarized in Table \ref{tbl:pathome_tbl1}.

\begin{figure}
    \centering
    \includegraphics[width = 0.9\textwidth]{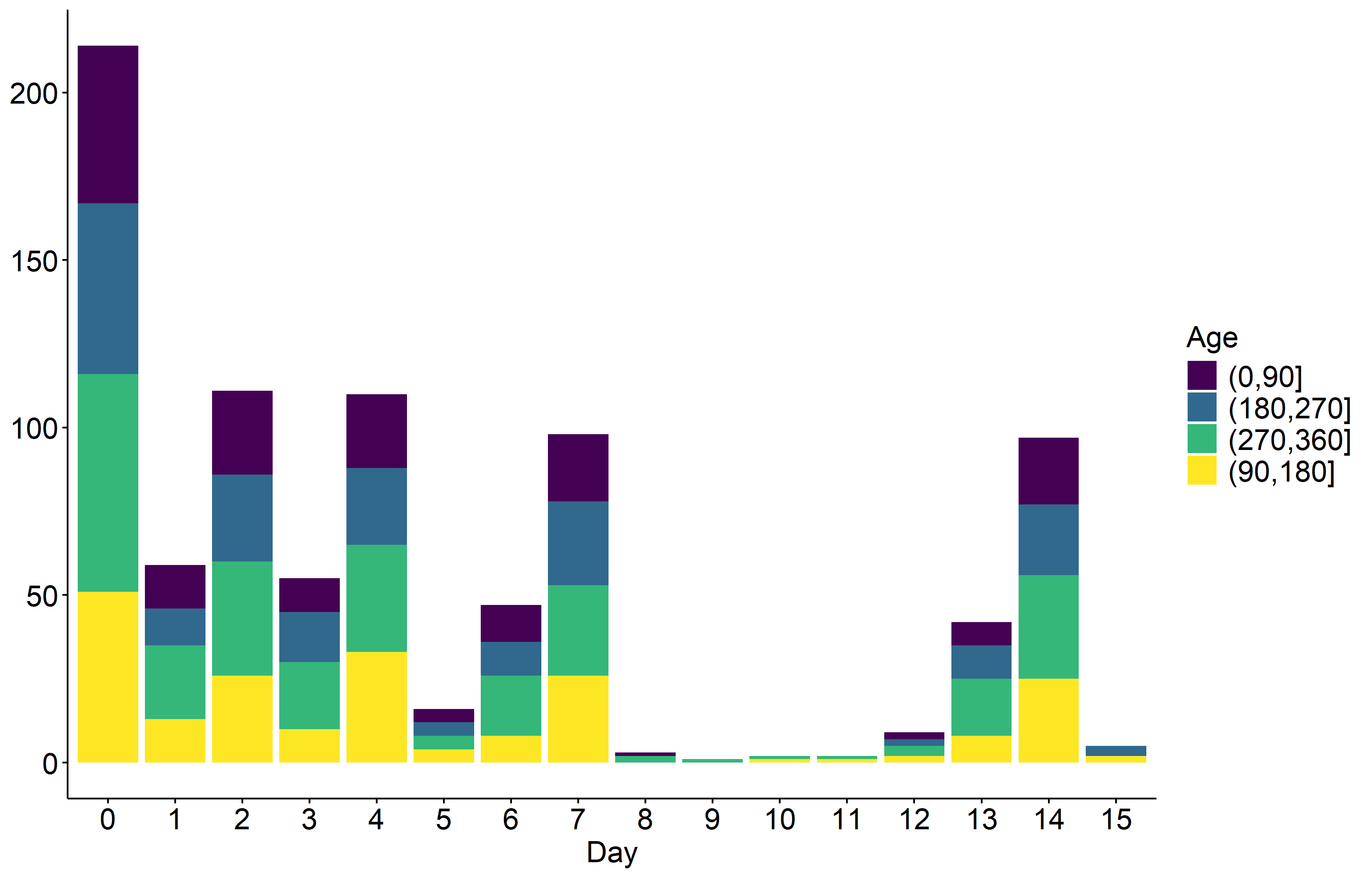}
    \caption{Number of infant diapers collected by day, broken down by age}
    \label{fig:day_counts}
\end{figure}

For each infant, diapers were provided to the caregivers in order to later collect stool samples on days 1, 3, 5, 7, and 14.  If a stool was unavailable on a prespecified day, field staff would return the subsequent day to attempt diaper collection.  The average (standard deviation) number of stool samples collected per child was 4.1 (1.2), and the counts of diapers collected on each day since enrollment is given in Figure \ref{fig:day_counts}, stressing the importance of methods that can accommodate varying time lags between observations. Each stool sample was analyzed by quantitative molecular detection methods targeting unique pathogen-specific gene sequences. 
While pathogen presence/absence was assigned for 19 pathogens, most were too sparse to be used in our analyses given our sample size of 214 households.  The pathogens included in this analysis were Enteroaggregative \textit{E. coli} (EAEC), Enterotoxigenic \textit{E. coli} (ETEC), typical enteropathogenic \textit{E. coli} (tEPEC), atypical enteropathogenic \textit{E. coli} (aEPEC), Shiga producing \textit{E. coli} (STEC), \textit{Campylobacter jejuni} (C. jejuni), \textit{Salmonella}, and \textit{Shigella}.  The numbers of infections present at baseline, new infections (as defined to be an absence at baseline with a positive detection at some later follow up), and no infections during study interval are given in Table \ref{tbl:pathome_pathogens}. 

We fit the DARE model to each of the eight pathogens, and subsequently applied the SUBSET method described in Section \ref{subsec:subset} to leverage information across pathogens, selecting the amount of shrinkage based on Bayes factors. We limited shrinkage of the regression coefficients relating to animals to those for which animals have been shown to act as a vector, namely \textit{aEPEC} \citep[e.g., ][]{krause2005investigation}, \textit{STEC} \citep[e.g., ][]{COBELJIC2005prevalence}, \textit{ETEC} \citep[e.g., ][]{daniel2016animal}, \textit{C. jejuni} \cite[e.g., ][]{manser1985survey}, and \textit{Salmonella} \citep[e.g., ][]{oloya2007evaluation}. As infants of various ages are likely to experience different exposures, we disaggregated infants into 3-month age groups: 0-3, 4-6, 7-9, and 10-12 months of age.

\begin{table}[h]
    \centering
    \begin{tabular}{lllll}
        \toprule
        Age   & (0,90] & (90,180] & (180,270] & (270,360]\\
        \midrule
        Number of infants  & 47 & 51 & 51 & 65\\
        \addlinespace
        City (\%) &&&& \\
        \hspace{3pc} Kisumu & 21 (44.7) & 20 (39.2) & 19 (37.3) & 22 (33.8)\\
        \hspace{3pc} Nairobi & 26 (55.3) & 31 (60.8) & 32 (62.7) & 43 (66.2)\\
        SES (\%) &&&& \\
        \hspace{3pc} Lower class & 24 (51.1) & 21 (41.2) & 24 (47.1) & 35 (53.8)\\
        \hspace{3pc}  Middle class & 23 (48.9) & 30 (58.8) & 27 (52.9) & 30 (46.2)\\
        Flood (\%) &&&&\\ 
        \hspace{3pc} No & 42 (89.4) & 43 (84.3) & 42 (82.4) & 56 (86.2)\\
        \addlinespace
         \hspace{3pc} Yes & 5 (10.6)& 8 (15.7) & 9 (17.6) & 9 (13.8) \\
        Household owns animals  (\%)  &&&&\\ 
        \hspace{3pc} No  & 42 (89.4) & 41 (80.4) & 45 (88.2) & 54 (83.1)\\
        \hspace{3pc} Yes & 5 (10.6) & 10 (19.6) & 6 (11.8) & 11 (16.9)\\
        Neighborhood animals enter compound (\%) &&&&\\
        \hspace{3pc} No & 33 (70.2) & 33 (64.7) & 36 (70.6) & 49 (75.4)\\
        \hspace{3pc} Yes & 14 (29.8) & 18 (35.3) & 15 (29.4) & 16 (24.6)\\
        Latrine (\%) &&&&\\
        \hspace{3pc} Private  & 35 (74.5) & 40 (78.4) & 37 (72.5) & 48 (73.8)\\
        \addlinespace
        \hspace{3pc} Public & 12 (25.5) & 11 (21.6) & 14 (27.5) & 17 (26.2)\\
        \bottomrule
    \end{tabular}
    \caption{Summary statistics for infants enrolled in the PATHOME study.}
    \label{tbl:pathome_tbl1}
\end{table}

\begin{table}[ht]
    \centering
    \begin{tabular}{lrrr}
        \toprule
        Pathogen & Present at baseline & New infection & No infection\\
        \midrule
        \textit{EAEC} & 107 & 54 & 53\\
        \textit{ETEC} & 35 & 33 & 146\\
        \textit{tEPEC} & 14 & 19 & 181\\
        \textit{aEPEC} & 41 & 38 & 135\\
        \textit{STEC} & 15 & 27 & 172\\
        \textit{Campylobacter jejuni} & 18 & 16 & 180\\
        \textit{Salmonella} & 39 & 37 & 138\\
        \textit{Shigella} & 60 & 29 & 125\\
        \bottomrule
    \end{tabular}
    \caption{Pathogen detection summary for infants enrolled in the PATHOME study.}
    \label{tbl:pathome_pathogens}
\end{table}

Figure \ref{fig:pathome} shows the point estimates and credible intervals for the dose accrual rate ratios. Only animal ownership displayed statistical significance, and interestingly this occurred around the time infants often learn to crawl.  For \textit{ETEC}, \textit{aEPEC}, \textit{STEC}, \textit{C. jejuni}, and \textit{Salmonella}, we estimated the dose accrual rate to be 5.3, 4.6, 6.0, 5.0, and 5.0 times higher respectively for those whose household owned animals compared to those whose household didn't.  The posterior probabilities that these rate ratios were greater than one exceeded 0.99 for these five pathogens.  

\begin{figure}[p]
    \centering
    \begin{subfigure}[b]{0.45\linewidth}
        \includegraphics[width = \textwidth]{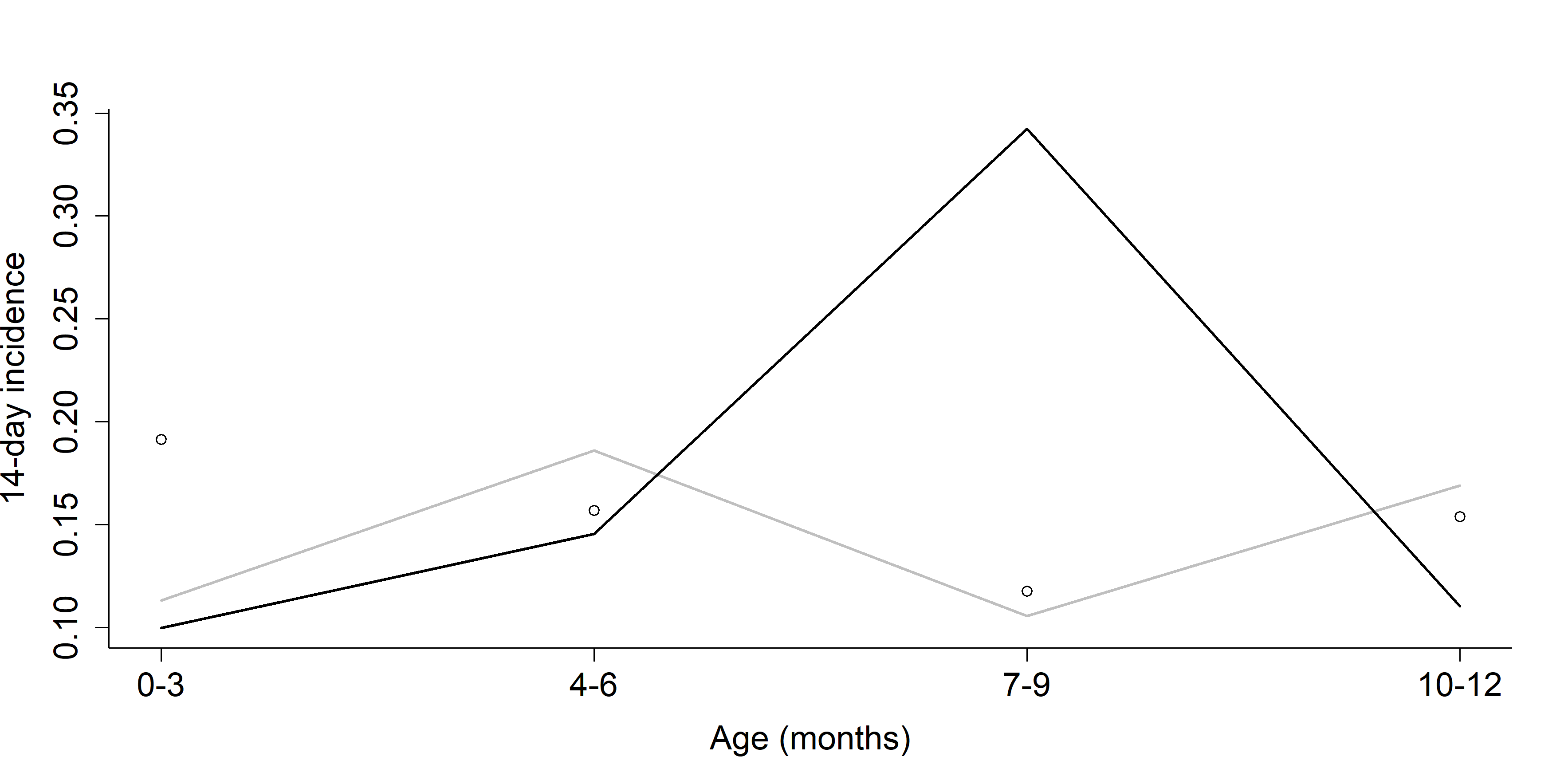}
        \caption{\textit{ETEC}}
    \end{subfigure}
    \begin{subfigure}[b]{0.45\linewidth}
        \includegraphics[width = \textwidth]{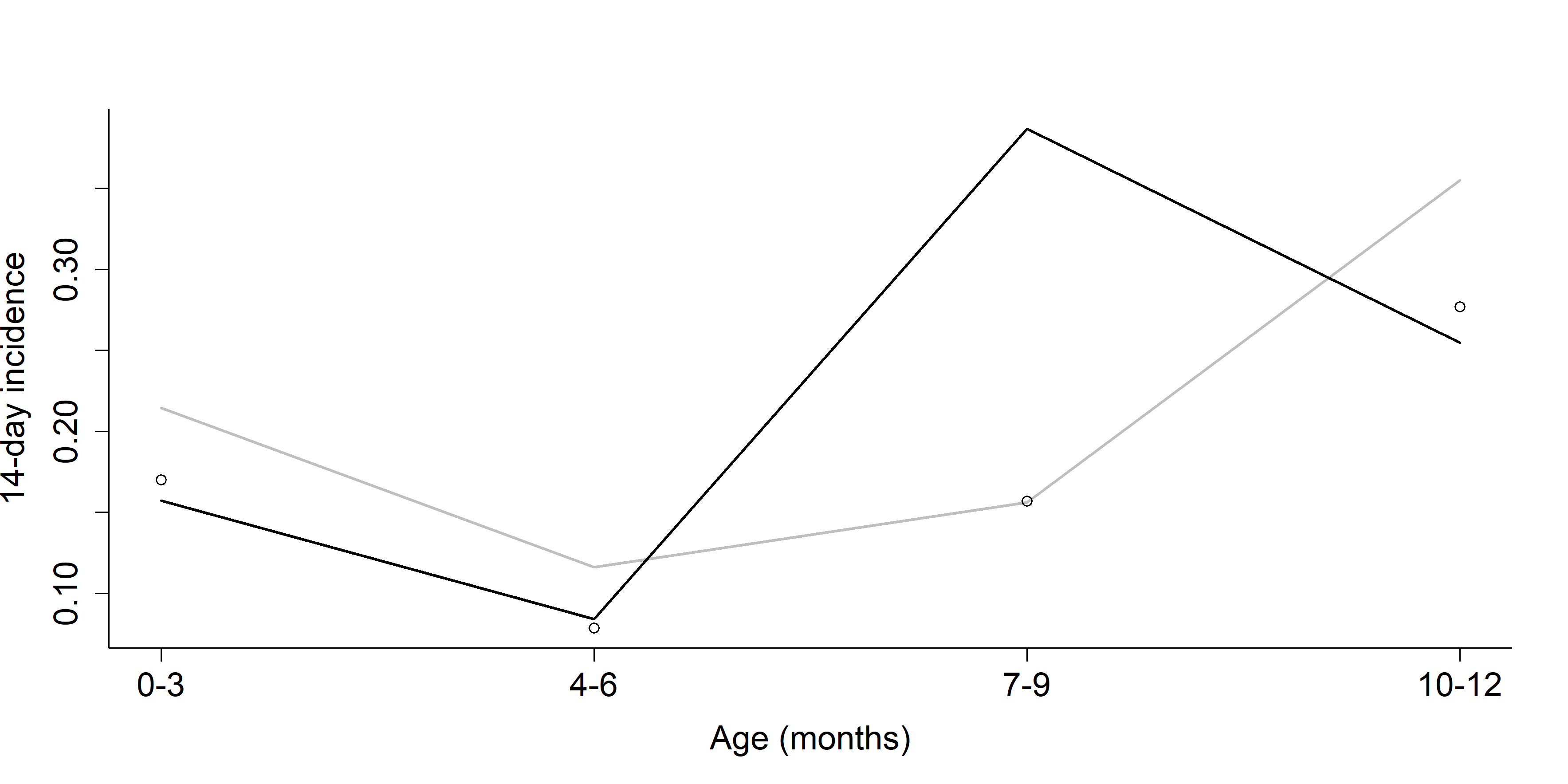}
        \caption{\textit{aEPEC}}
    \end{subfigure} \\

    \begin{subfigure}[b]{0.45\linewidth}
        \includegraphics[width = \textwidth]{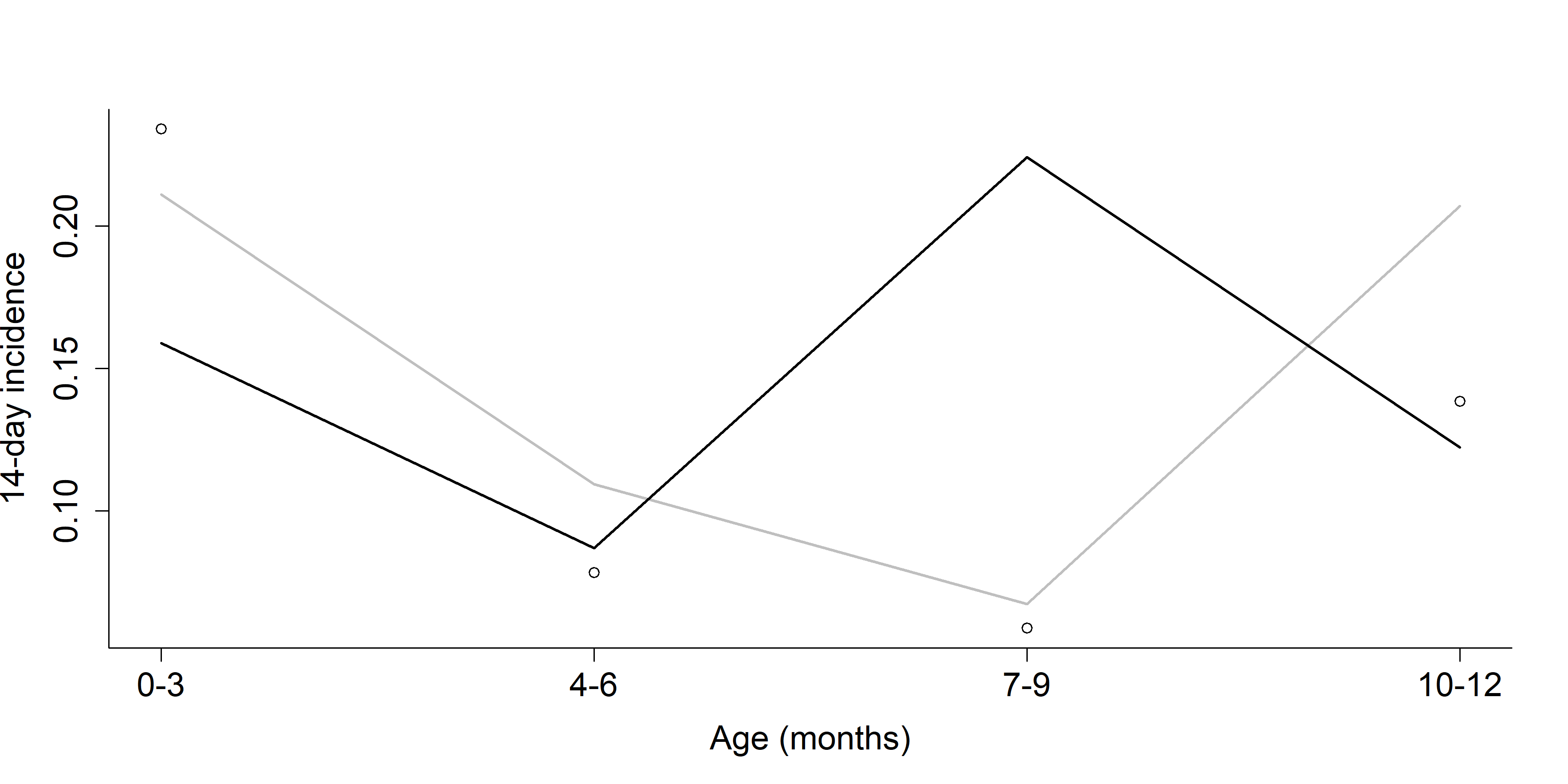}
        \caption{\textit{STEC}}
    \end{subfigure}
    \begin{subfigure}[b]{0.45\linewidth}
        \includegraphics[width = \textwidth]{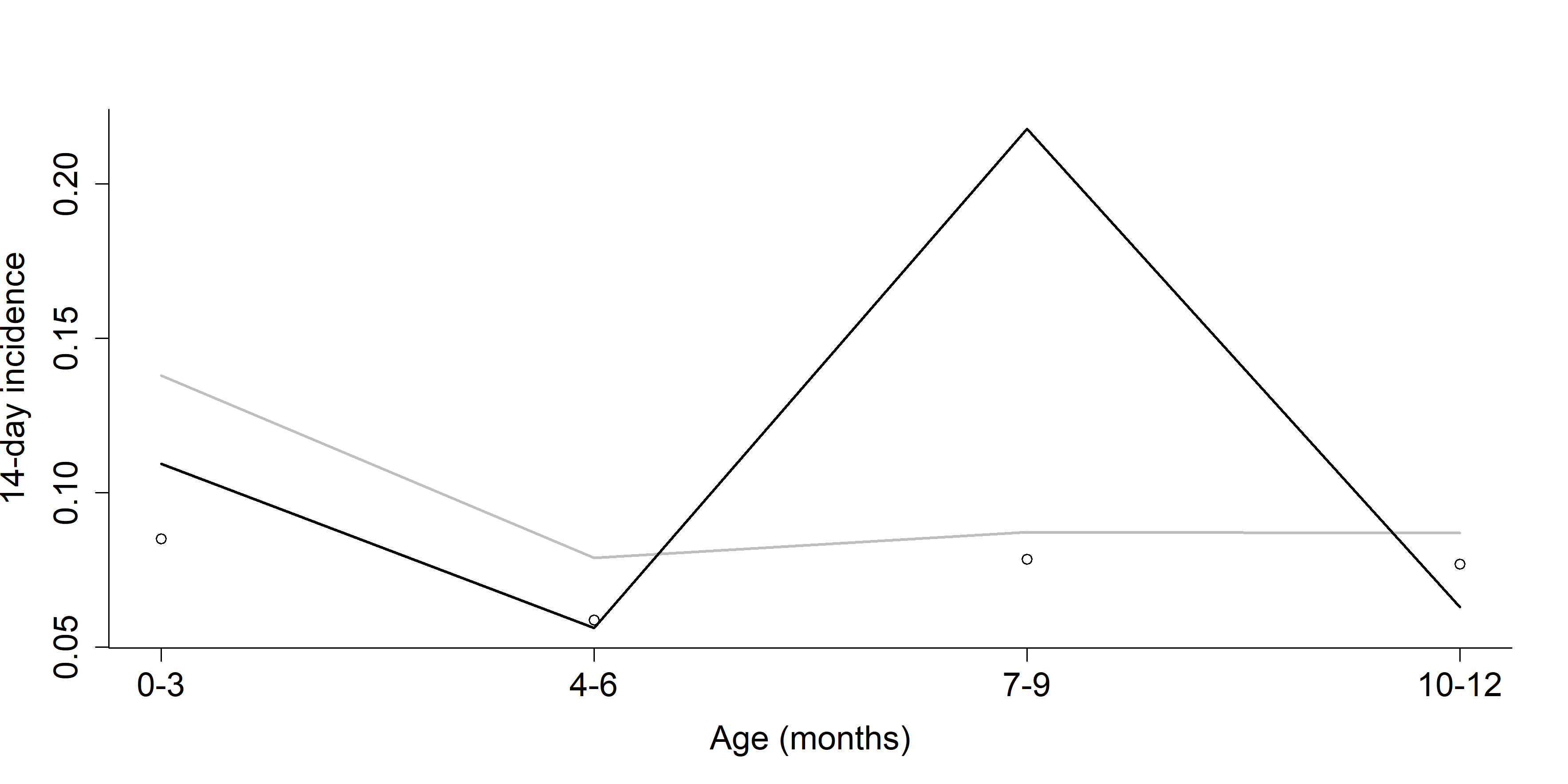}
        \caption{\textit{C. jejuni}}
    \end{subfigure} \\

    \begin{subfigure}[b]{0.45\linewidth}
        \includegraphics[width = \textwidth]{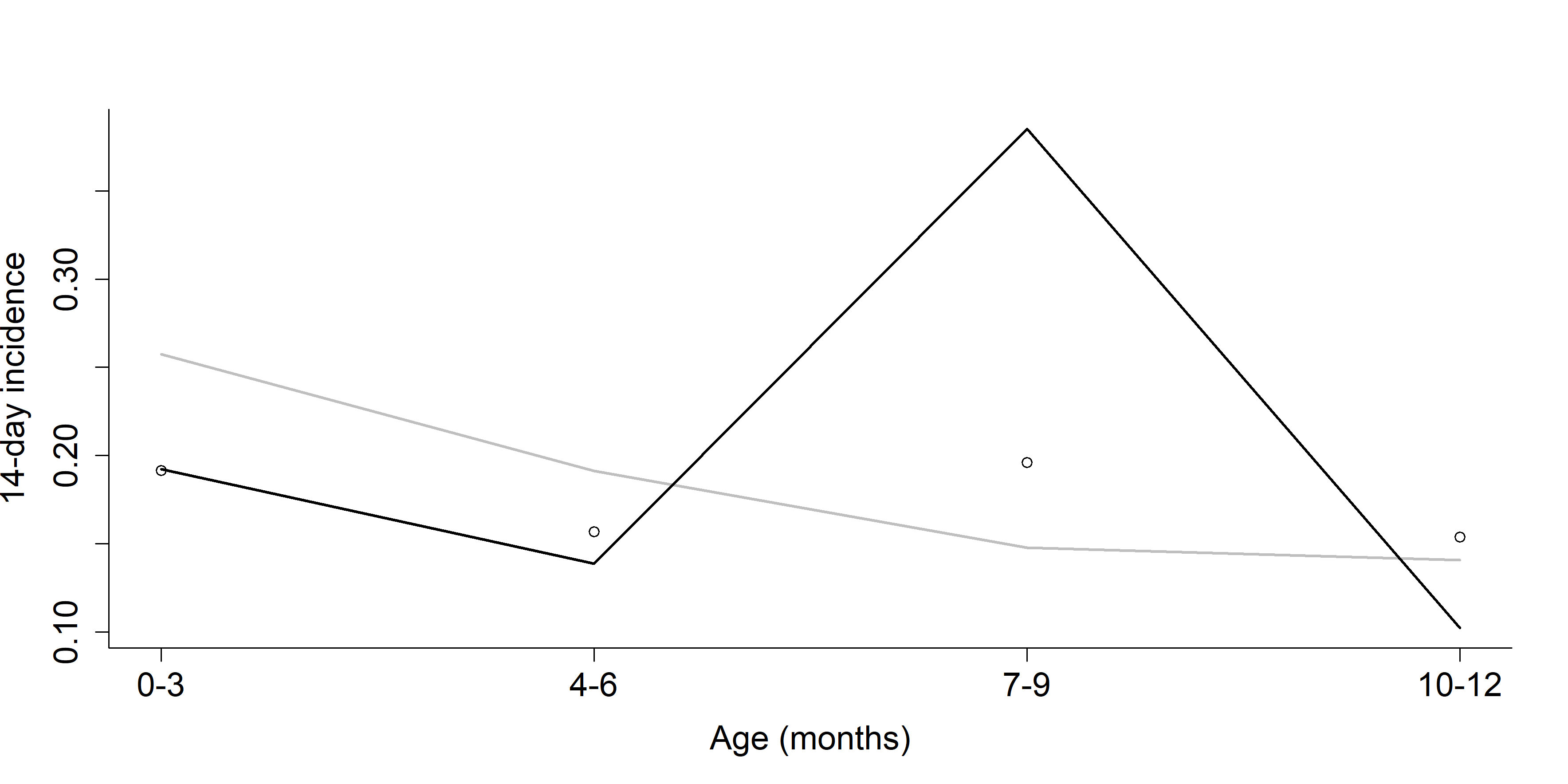}
        \caption{\textit{Salmonella}}
    \end{subfigure} 
    \begin{subfigure}[b]{0.45\linewidth}
        \includegraphics[width = \textwidth]{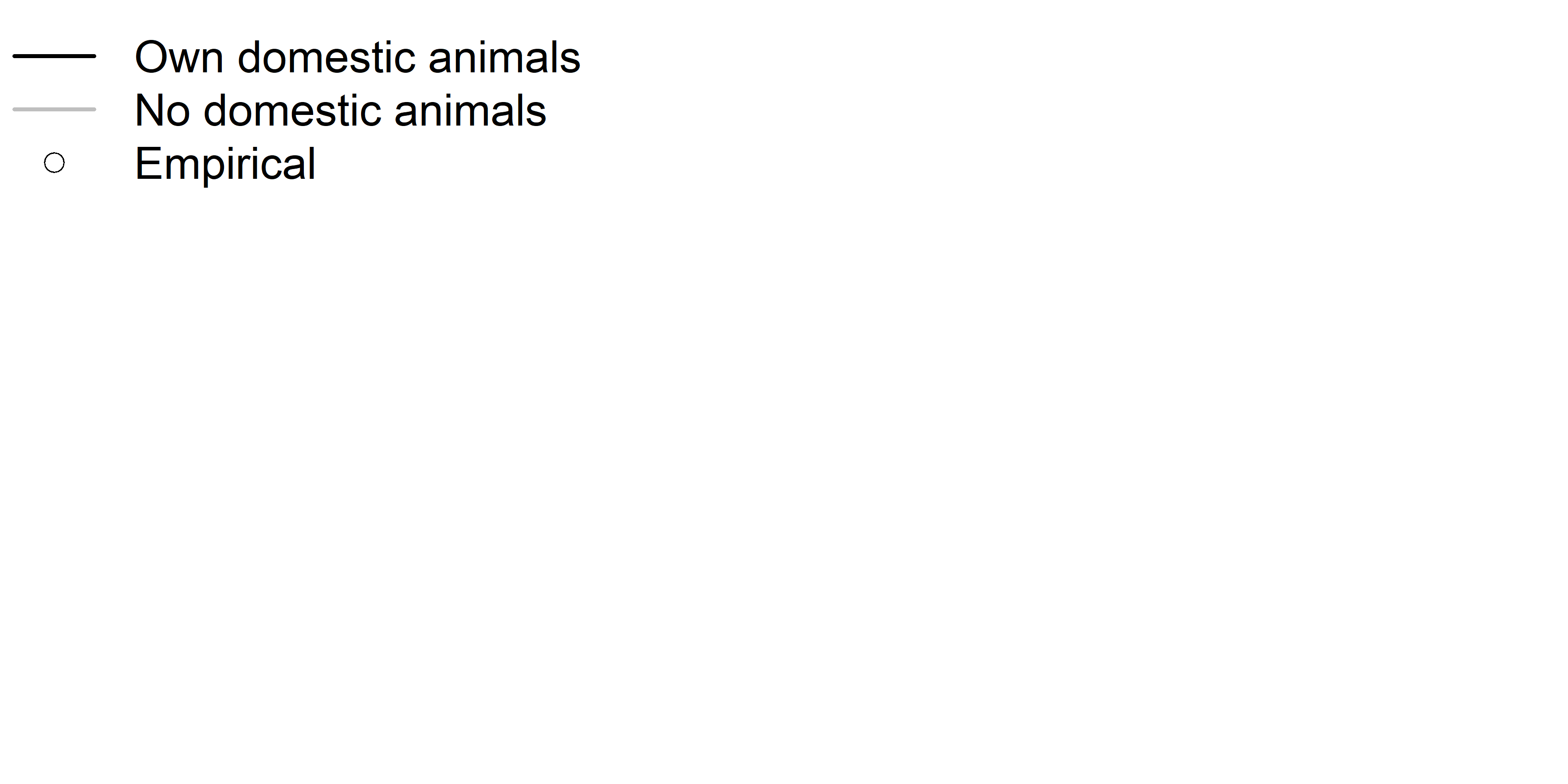}
    \end{subfigure} 
    
    \caption{Estimated 14-day incidence proportion by domestic animal ownership, holding all other covariates fixed at their modal value}
    \label{fig:pathome-incidence}
\end{figure}

We also examined the effect of animal ownership on incidence proportion for these five pathogens, holding all other covariates fixed at their modal value; these values are given in Figure \ref{fig:pathome-incidence}.  For \textit{ETEC}, \textit{aEPEC}, \textit{STEC}, \textit{C. jejuni}, and \textit{Salmonella} respectively, the incidence proportion rate for 7-9 month olds was estimated to be 3.2, 2.5, 3.3, 2.5, and 2.6 times higher for those whose households owned animals compared to those whose households did not.

\begin{figure}[p]
    \centering
    \begin{subfigure}[b]{0.5\linewidth}
        \includegraphics[width=\textwidth]{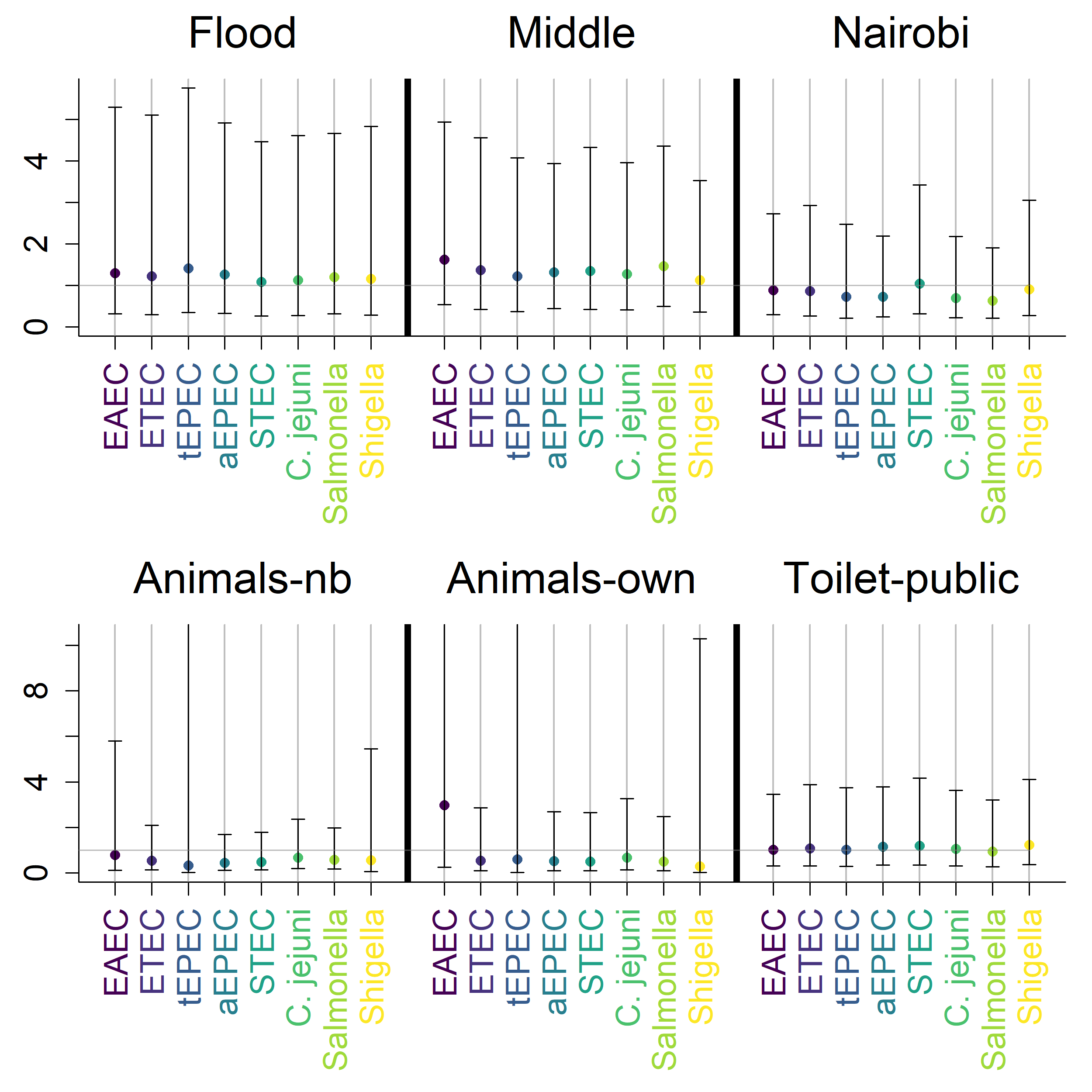}
        \caption{0-3 months}
    \end{subfigure} \\
    
    \begin{subfigure}[b]{0.5\linewidth}
        \includegraphics[width=\textwidth]{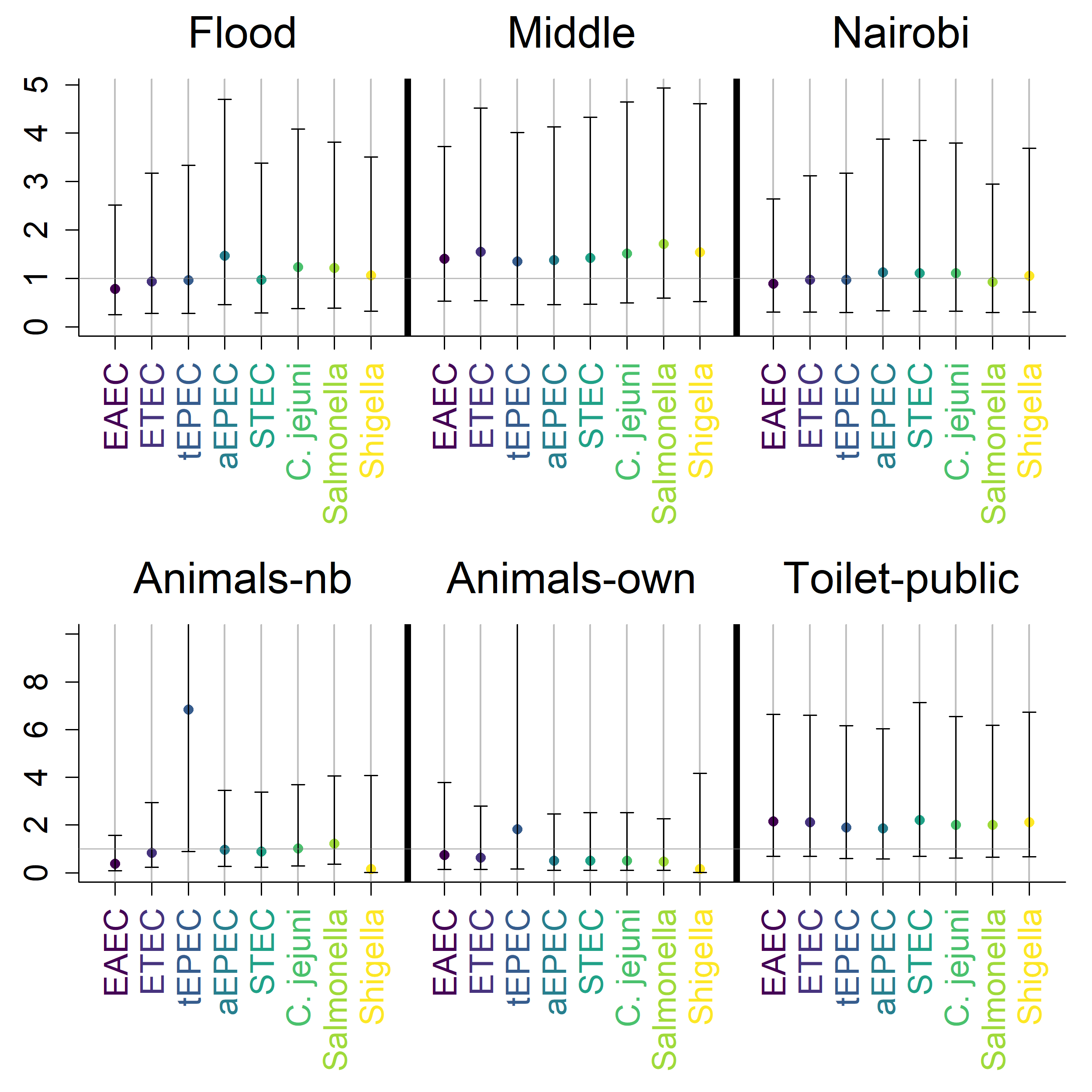}
        \caption{4-6 months}
    \end{subfigure}
    
    \caption{Dose accrual rate ratios for infants living in Kisumu and Nairobi, Kenya. The reference categories for ``Middle'', ``Nairobi'', and ``Toilet-public'' are low-class neighborhood, Kisumu, and access to a private toilet respectively.  ``Animals-nb'' refers to whether neighborhood animals enter the compound, and ``Animals-own'' refers to whether the household owns domestic animals.}
    \label{fig:pathome}
\end{figure}
\begin{figure}[p]
    \ContinuedFloat
    \centering
    \begin{subfigure}[b]{0.5\linewidth}
        \includegraphics[width=\textwidth]{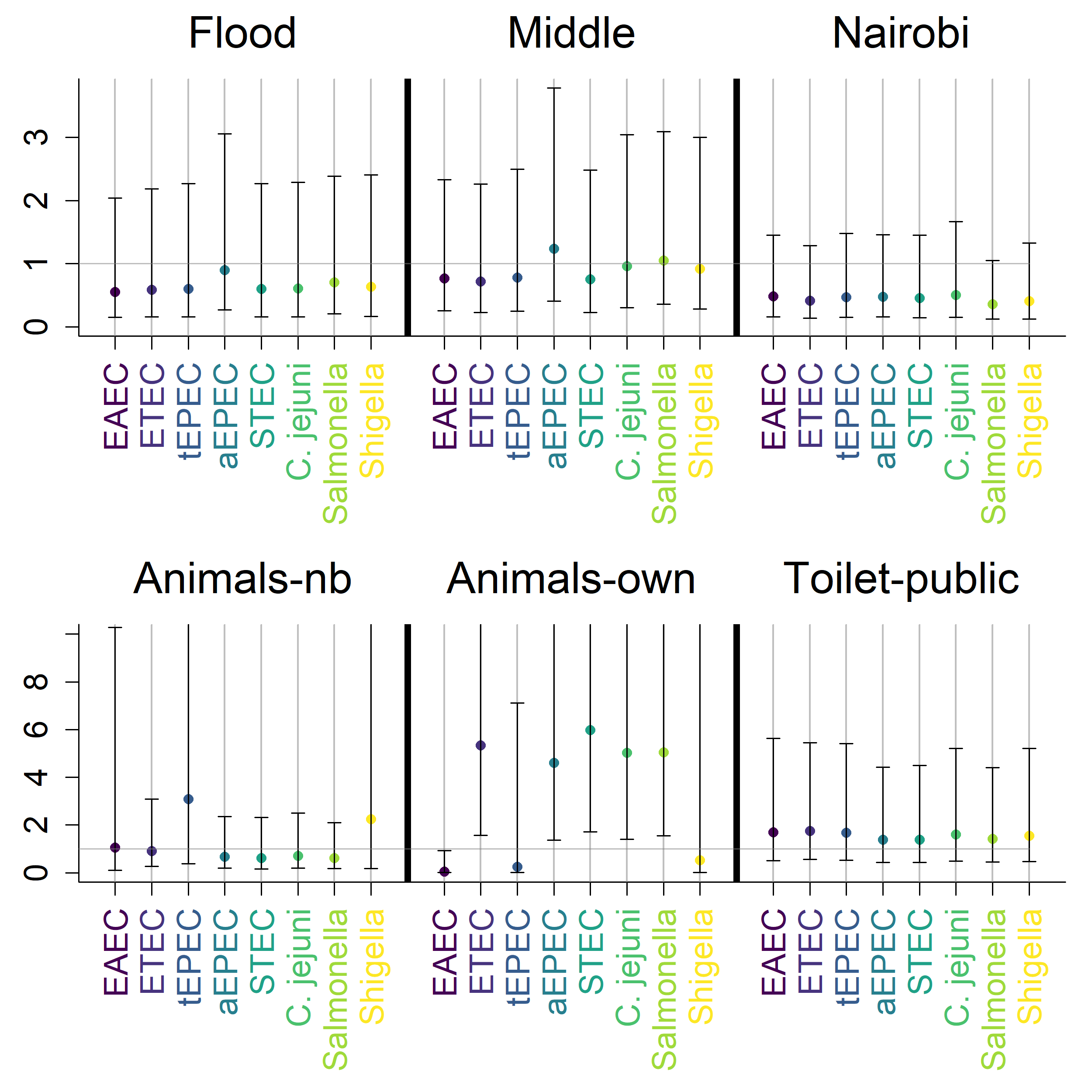}
        \caption{7-9 months}
    \end{subfigure} \\
    
    \begin{subfigure}[b]{0.5\linewidth}
        \includegraphics[width=\textwidth]{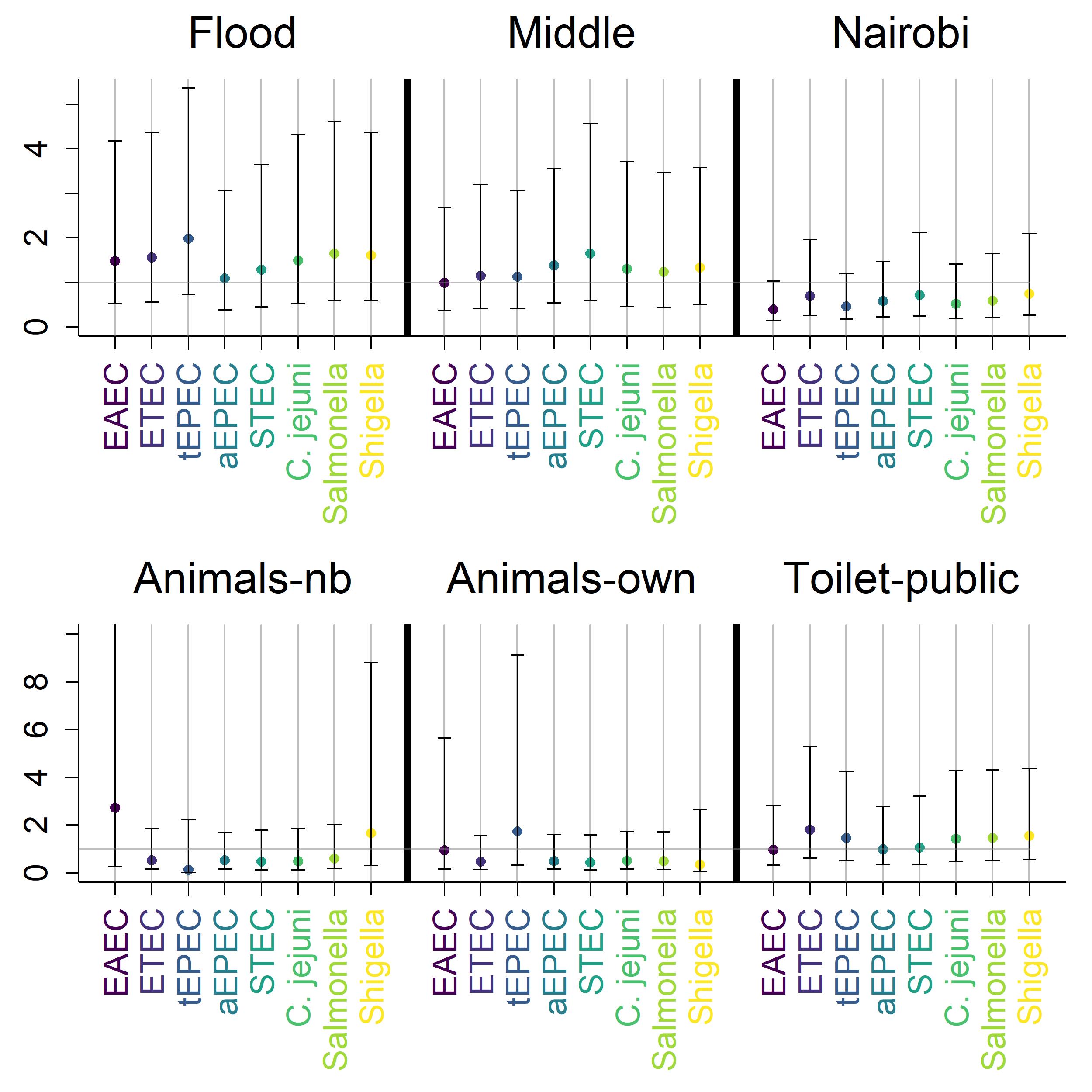}
        \caption{10-12 months}
    \end{subfigure}
    
    \caption{(Cont'd) Dose accrual rate ratios for infants living in Kisumu and Nairobi, Kenya. The reference categories for ``Middle'', ``Nairobi'', and ``Toilet-public'' are low-class neighborhood, Kisumu, and access to a private toilet respectively.  ``Animals-nb'' refers to whether neighborhood animals enter the compound, and ``Animals-own'' refers to whether the household owns domestic animals.}
    \label{fig:pathome_contd}
\end{figure}

\section{Discussion}
\label{sec:discussion}
Enteric disease is a significant source of morbidity and mortality globally, especially in children living in low- to middle-income countries.  Enteric diseases occur through the ingestion of pathogens, and understanding how various risk factors affect dose accrual rates is vital to developing effective interventions.  We have proposed a novel approach to estimating the effect individual-level characteristics have on pathogenic dose accrual rates.  Our approach provides a biologically plausible infection model that attempts to account for the four sources of variability we have described- risk factors, dose density, number of pathogens ingested, and pathogen survival rate (see Figure \ref{fig:sources_of_variability}).

Our approach allows measurements taken at unevenly spaced time points, can flexibly handle various disease-appropriate dose response models, and our simulation study suggests that the coverage rate is maintained at or very nearly at the nominal rate even under misspecification of the dose-response model.  We have further provided a method for leveraging information across multiple pathogens within a single study. 

We anticipate in most cases it will be clear if a risk factor may have an effect on dose accrual rates vs. within-host pathogen survival probability. In such cases, parameter interpretability of DARE is a strength, as the exponentiated regression coefficients provide the dose accrual rate ratio corresponding to a unit increase in the covariate of interest, holding all other covariates constant.  However, interpretability is a limitation of our proposed approach in cases where the risk factor may affect both the dose accrual and the dose survival rate.  There is perfect confounding in this situation.  Intuitively this makes sense and appears to be unavoidable, since without exceedingly granular microbiological data collection procedures, it is not possible to disambiguate the number of pathogens ingested from the pathogens' survival rates, when both survival and dose accrual depend on the same factor(s).  It is, however, still possible to understand the general pattern (positive or negative) in the incidence rate due to such a risk factor, even if the precise cause is unknown. Our proposed methods can be implemented in the R programming language \citep{rlanguage} via the R package found at \url{https://github.com/dksewell/dare}.

\bibliographystyle{WileyNJD-AMS} 
\bibliography{SewellBibFiles}

\end{document}